\documentclass{article}
\usepackage{amssymb}
\usepackage{graphicx}

\input{tcilatex}

\begin{document}

\title{Quantum Stackelberg duopoly in noninertial frame}
\author{Salman Khan\thanks{%
sksafi@phys.qau.edu.pk}, M. Khalid Khan \\
Department of Physics, Quaid-i-Azam University, \\
Islamabad 45320, Pakistan.}
\maketitle

\begin{abstract}
We study the influence of Unruh effect on quantum Stackelberg duopoly. We
show that the acceleration of noninertial frame strongly effects the payoffs
of the firms. The validation of the subgame perfect Nash equilibrium is
limited to a particular range of acceleration of the noninertial frame. The
benefit of initial state entanglement in the quantum form of the duopoly in
inertial frame is adversely affected by the acceleration. The duopoly can
become as a follower advantage only in a small region of the acceleration.%
\end{abstract}

Game theory is the mathematical study of interaction among independent, self
interested agents. It emerged from the work of Von Neumann \cite{von neumann}%
, and is now used in various disciplines like economics, biology, medical
sciences, social sciences and physics \cite{Piotrowski,Baaquie}. Due to
dramatic development in quantum information theory \cite{Neilson}, the game
theorists [5-9] have made strenuous efforts to extend the classical game
theory into the quantum domain. The first attempt in this direction was made
by Meyer \cite{Meyer} by quantizing a simple coin tossing game. Applications
of quantum games are reviewed by several authors \cite{Taksu1,Tsutsui1}. A
formulation of quantum game theory based on the Schmidt decomposition is
presented by Ichikawa et al. \cite{Taksu2}.

In quantum games, results different from the classical counterparts are
obtained by using the fascinating feature of quantum mechanics "the
entanglement". Recently, the study of quantum entanglement of various fields
has been extended to the relativistic setup \cite{Alsing,Ling,Gingrich,Pan,
Schuller, Terashima} and interesting results about the behavior of
entanglement have been obtained. Alsing \textit{et al} \cite{Alsing} have
shown that the entanglement between two modes of a free Dirac field is
degraded by the Unruh effect and asymptotically reaches a nonvanishing
minimum value in the infinite acceleration.

In this letter, we study the influence of Unruh effect on the payoffs
function of the firms in the quantum Stackelberg duopoly. We show that the
payoffs function of the firms are strongly influenced by the acceleration of
the noninertial frame. It is shown that for small values of acceleration the
duopoly is leader advantage and it becomes the follower advantage in the
range of large values of acceleration. Unlike the quantum form of the
duopoly in inertial frames, the benefit of initial state entanglement is
adversely affected in the noninertial frames. We show that for a maximally
entangled initial state, the Unruh effect damps the payoffs considerably as
compared to the case of unentangled initial state. Furthermore, it is shown
that the Unruh effect limits the validation of the subgame perfect Nash
equilibrium outcome to a particular range of values of the acceleration of
the frame. The payoffs of the firms vanish, irrespective of the initial
state entanglement, at a particular value of the acceleration.

The Stackelberg duopoly is a market game, which is a modified form of the
Cournot duopoly. In the Cournot duopoly, two firms simultaneously put a
homogeneous product into a market and guess that what action the opponent
will take. The Stackelberg duopoly is a dynamic model of duopoly in which
one firm, say firm $A$, moves first and the other firm, say $B$, goes after.
Before making its decision, firm $B$ observes the move of firm $A$. This
transforms the static nature of the Cournot duopoly to a dynamic one. Firm $%
A $ is usually called the leader and firm $B$ the follower, on this basis
the game is also called the leader-follower model \cite{Gibbons}. In the
classical Stackelberg duopoly, it is assumed that firm $B$ will respond
optimally to the strategic decision of firm $A$. As firm $A$ can precisely
predict firm $B$'s strategic decision, firm $A$ chooses its move in such a
way that maximizes its own payoff. This informational asymmetry makes the
Stackelberg duopoly as the first mover advantage game. The quantum
Stackelberg duopoly has been studied under various circumstances and
interesting results have been obtained \cite{Iqbal2,Xia1,Xia2,salman}

We consider two firms, $A$ and $B$, that share an entangled initial state of
two qubits at a point in flat Minkowski spacetime. Then firm $B$ moves with
a uniform acceleration and firm $A$ stays stationary. Let the two modes of
Minkowski spacetime that correspond to firm $A$ and firm $B$ are,
respectively, given by $|n\rangle _{A}$ and $|n\rangle _{B}$. We assume that
the firms share the following entangled initial state%
\begin{equation}
|\psi _{i}\rangle =\cos \theta |00\rangle _{A,B}+\sin \theta |11\rangle
_{A,B}  \label{1}
\end{equation}%
\begin{figure}[h]
\begin{center}
\begin{tabular}{ccc}
\vspace{-0.5cm} \includegraphics[scale=0.4]{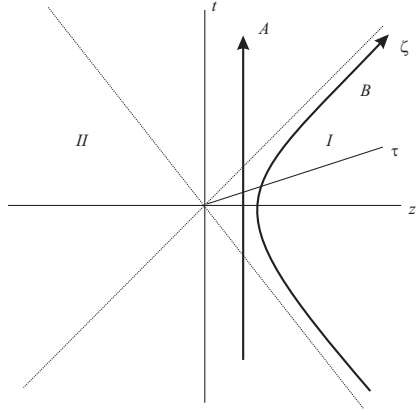}\put(-220,220) &  & 
\end{tabular}%
\end{center}
\caption{Rindler spacetime diagram: A uniformly accelerated observer B (firm
B) moves on a hyperbola with acceleration $a$ in region $I$ and is causally
disconnected from region $II$.}
\label{Fig 1}
\end{figure}
where $\theta $ is a measure of entanglement. The state is maximally
entangled at $\theta =\frac{\pi }{4}$. The first entry in each ket of Eq. (%
\ref{1}) corresponds to firm $A$ and the second entry corresponds to firm $B$%
. From the accelerated firm $B$'s frame, the Minkowski vacuum state is found
to be a two-mode squeezed state \cite{Alsing}%
\begin{equation}
|0\rangle _{M}=\cos r|0\rangle _{I}|0\rangle _{II}+\sin r|1\rangle
_{I}|1\rangle _{II},  \label{2}
\end{equation}%
where $\cos r=\left( e^{-2\pi \omega c/a}+1\right) ^{-1/2}$. The constant $%
\omega $, $c$ and $a$, in the exponential stand, respectively, for Dirac
particle's frequency, light's speed in vacuum and firm $B$'s acceleration.
In Eq. (\ref{2}) the subscripts $I$ and $II$ of the kets represent the
Rindler modes in region $I$ and $II$, respectively, in the Rindler spacetime
diagram (see Fig. ($1$)). The excited state in Minkowski spacetime is
related to Rindler modes as follow \cite{Alsing}%
\begin{equation}
|1\rangle _{M}=|1\rangle _{I}|0\rangle _{II}.  \label{3}
\end{equation}

In terms of Minkowski modes for firm $A$ and Rindler modes for firm $B$, the
entangled initial state of Eq. (\ref{1}) by using Eqs. (\ref{2}) and (\ref{3}%
) becomes%
\begin{equation}
|\psi \rangle _{A,I,II}=\cos \theta \cos r|0\rangle _{A}|0\rangle
_{I}|0\rangle _{II}+\cos \theta \sin r|0\rangle _{A}|1\rangle _{I}|1\rangle
_{II}+\sin \theta |1\rangle _{A}|1\rangle _{I}|0\rangle _{II}.  \label{4}
\end{equation}%
Since firm $B$ is causally disconnected from region $II$, we must take trace
over all the modes in region $II$. This leaves the following density matrix
between the two firms,%
\begin{equation}
\rho _{A,I}=\left( 
\begin{array}{cccc}
\cos ^{2}r\cos ^{2}\theta & 0 & 0 & \cos r\cos \theta \sin \theta \\ 
0 & \cos ^{2}\theta \sin ^{2}r & 0 & 0 \\ 
0 & 0 & 0 & 0 \\ 
\cos r\cos \theta \sin \theta & 0 & 0 & \sin ^{2}\theta%
\end{array}%
\right) .  \label{5}
\end{equation}

In the quantum Stackelberg duopoly, each firm has two possible strategies $I$%
, the identity operator and $C$, the inversion operator or Pauli's bit-flip
operator. Let $x$ and $1-x$ stand for the probabilities of $I$ and $C$ that
firm $A$ applies and $y$, $1-y$, are the probabilities that firm $B$
applies, respectively. The final density matrix is given by \cite%
{Marrinatto2}%
\begin{eqnarray}
\rho _{f} &=&xyI_{A}\otimes I_{B}\ \rho _{A,I}\ I_{A}^{\dag }\otimes
I_{B}^{\dag }+x\left( 1-y\right) I_{A}\otimes C_{B}\ \rho _{A,I}\
I_{A}^{\dag }\otimes C_{B}^{\dag }  \nonumber \\
&&+y\left( 1-x\right) C_{A}\otimes I_{B}\ \rho _{A,I}\ C_{A}^{\dag }\otimes
I_{B}^{\dag }  \nonumber \\
&&+\left( 1-x\right) \left( 1-y\right) C_{A}\otimes C_{B}\ \rho _{A,I}\
C_{A}^{\dag }\otimes C_{B}^{\dag },  \label{6}
\end{eqnarray}%
where $\rho _{A,I}$ is the density matrix given by Eq. (\ref{5}).

Suppose that the players' moves in the quantum Stackelberg duopoly are given
by probabilities lying in the range $[0,1]$. In the classical form of the
duopoly, the moves of firms $A$ and $B$ are given by quantities $q_{1}$ and $%
q_{2}$, which have values in the range $[0,\infty )$. We assume that firms $%
A $ and $B$ agree on a function that uniquely defines a real positive number
in the range $(0,1]$ for every quantity $q_{1}$, $q_{2}$ in $[0,\infty )$.
Such a function is given by $1/(1+q_{i})$, so that firms $A$ and $B$ find $x$
and $y$, respectively, as 
\begin{equation}
x=\frac{1}{1+q_{1}}\mathrm{,\qquad }y=\frac{1}{1+q_{2}}  \label{7}
\end{equation}%
The payoffs of firms $A$ and $B$ are given by the following trace operations%
\begin{equation}
P_{A}\left( q_{1},q_{2}\right) =\mathrm{Tr}\left[ \rho _{f}P_{A}^{\mathrm{op}%
}\left( q_{1},q_{2}\right) \right] \mathrm{,\qquad }P_{B}\left(
q_{1},q_{2}\right) =\mathrm{Tr}\left[ \rho _{f}P_{B}^{\mathrm{op}}\left(
q_{1},q_{2}\right) \right] ,  \label{8}
\end{equation}%
where $P_{A}^{\mathrm{op}}$, $P_{B}^{\mathrm{op}}$ are payoff operators of
the firms and are given by%
\begin{eqnarray}
P_{A}^{\mathrm{op}}\left( q_{1},q_{2}\right) &=&\frac{q_{1}}{q_{12}}\left(
k\rho _{11}-\rho _{22}-\rho _{33}\right) ,  \nonumber \\
P_{B}^{\mathrm{op}}\left( q_{1},q_{2}\right) &=&\frac{q_{2}}{q_{12}}\left(
k\rho _{11}-\rho _{22}-\rho _{33}\right) ,  \label{9}
\end{eqnarray}%
where $\rho _{ii}$ are the diagonal elements of the final density matrix, $k$
is a constant as given in Ref. \cite{Gibbons} and $q_{12}$ is given by%
\begin{equation}
q_{12}=\frac{1}{\left( 1+q_{1}\right) \left( 1+q_{2}\right) }.  \label{10}
\end{equation}

The backward-induction outcome in the Stackelberg duopoly is found by first
finding the reaction function $R_{2}\left( q_{1}\right) $ of firm $B$ to an
arbitrary quantity $q_{1}$ chosen by firm $A$. It is found by
differentiating firm $B$'s payoff with respect to $q_{2}$, and maximizing
the result for $q_{1}$ and can be written as%
\begin{equation}
R_{2}\left( q_{1}\right) =\max P_{B}\left( q_{1},q_{2}\right)  \label{11}
\end{equation}%
Once firm $B$ chooses this quantity, firm $A$ can compute its optimization
problem by differentiating its own payoff with respect to $q_{1}$ and then
maximizing it to find the value $q_{1}=q_{1}^{\ast }$. Using the value of $%
q_{1}^{\ast }$ in Eq. (\ref{11}), we can get the value of $q_{2}^{\ast }$.
These quantities define the backward-induction outcome of the quantum
Stackelberg duopoly and represent the subgame perfect Nash equilibrium. The
payoffs of the firms at the subgame perfect Nash equilibrium can be found
using Eq. (\ref{8}). 
\begin{figure}[h]
\begin{center}
\begin{tabular}{ccc}
\vspace{-0.5cm} \includegraphics[scale=0.8]{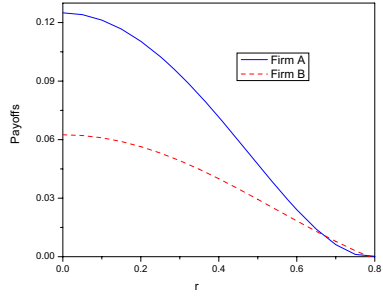}\put(-220,220) &  & 
\end{tabular}%
\end{center}
\caption{\textbf{(}color online) The payoffs are plotted at the subgame
perfect Nash equilibrium against the acceleration $r$ for unentangled
initial state. The value of $k$ is set to $1$. The solid line represents the
payoff of firm $A$ and the dotted line represents the Payoff of firm $B$.}
\label{Fig2}
\end{figure}

The subgame perfect Nash equilibrium outcome of the duopoly becomes%
\begin{eqnarray}
q_{1}^{\ast } &=&\frac{\cos ^{2}\theta (k\cos ^{2}r-\sin ^{2}r)}{2(\cos
^{2}r\cos ^{2}\theta +\sin ^{2}\theta )}  \nonumber \\
q_{2}^{\ast } &=&\frac{4\cos ^{2}\theta (k\cos ^{2}r-\sin ^{2}r)(\cos
^{2}r\cos ^{2}\theta +\sin ^{2}\theta )}{%
\begin{array}{c}
(3-k+12\cos 2r+(1+k)\cos 4r)\cos ^{4}\theta \\ 
-8\cos ^{2}\theta ((-4+k^{2})\cos ^{2}r+k\sin ^{2}r)\sin ^{2}\theta +16\sin
^{4}\theta%
\end{array}%
}  \label{15}
\end{eqnarray}%
It is important to note that the result of Eq. (\ref{15}) for unentangled
initial state ($\theta =0$) reduces to the classical result when we put the
acceleration $r=0$. Similarly the results of Ref. \cite{Iqbal2} for the
maximal entangled initial state are retrieved for $\theta =\pi /4$ and $r=0$%
. In the classical form of the duopoly the subgame perfect Nash equilibrium
is a point, whereas in this case, it is a function of both entanglement
angle $\theta $ and the acceleration $r$\ of firm $B$'s frame. The payoffs
of the firms at the subgame perfect Nash equilibrium for unentangled initial
state, when $k=1$, are given as%
\begin{eqnarray}
P_{A} &=&\frac{1}{8}\cos ^{2}2r\sec ^{2}r  \nonumber \\
P_{B} &=&\frac{\cos ^{2}r\cos 2r}{4(3+\cos 2r)}  \label{16}
\end{eqnarray}%
The payoffs of the firms for a maximally entangled initial state, with $k=1$%
, become%
\begin{eqnarray}
P_{A} &=&\frac{\cos ^{2}2r}{8(3+\cos 2r)}  \nonumber \\
P_{B} &=&\frac{\cos ^{2}2r(3+\cos 2r)\sec ^{2}r}{32(6+\cos 2r)}  \label{17}
\end{eqnarray}%
\begin{figure}[h]
\begin{center}
\begin{tabular}{ccc}
\vspace{-0.5cm} \includegraphics[scale=0.8]{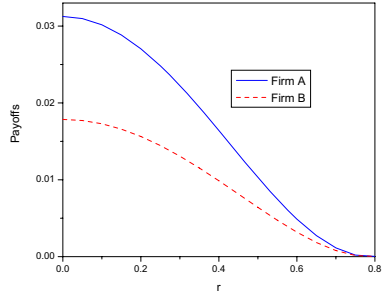}\put(-220,220) &  & 
\end{tabular}%
\end{center}
\caption{\textbf{(}color online) The payoffs are plotted at the subgame
perfect Nash equilibrium against the acceleration $r$ for maximally
entangled initial state. The value of $k$ is set to $1$. The solid line
represents the payoff of firm $A$ and the dotted line represents the Payoff
of firm $B$.}
\label{Fig3}
\end{figure}
The existence of the Nash equilibrium requires that the firms' moves ($%
q_{1}^{\ast }$ and $q_{2}^{\ast }$) should have positive values. It can
easily be checked from Eq. (\ref{15}) that for both unentangled and
maximally entangled initial states the move of firm $A$ becomes negative for 
$r\geq \pi /4$. Hence no Nash equilibrium exists for the values of $r$ at
which $q_{1}^{\ast }$ becomes negative. Thus the range of the acceleration
in which the acceleration parameter $r$ is given by $\pi /4\leq r\leq \pi /2$
is not a physically meaningful range for the Stackelberg duoply. To see how
the payoffs are influenced by the acceleration in its physically meaningful
range, we plot it against\ the acceleration parameter $r$. In Fig. $2$, we
show the plot of the firms' payoffs against $r$ for the unentangled initial
state. It can be seen that for smaller values of the acceleration, the
duopoly is leader advantage and the payoffs decrease with the increasing
value of the acceleration. At $r=0.66$ there happens a critical point at
which both firms are equally benefitted. From this point onward, the payoff
of firm $A$ rapidly decreases and becomes zero at $r=0.76$. The duopoly
becomes follower advantage in the region $0.66<r<0.78$. The payoff of the
follower firm reaches zero at $r=0.78$. The payoffs of the firms for the
maximally entangled initial state are plotted in Fig. $3$. It can be seen
that the payoffs of the firms are highly damped as compared to the case of
unentangled initial state\ and the duopoly is follower advantage for the
whole range of the acceleration in which the Nash equilibrium exists. The
payoffs of both firms becomes zero at $r=0.75$. In Fig. $4$, we plot the
payoffs of the firms against the entanglement angle $\theta $. It is seen
that the payoffs decrease with the increasing degree of entanglement in the
initial state. The duopoly is follower advantage for smaller value of $%
\theta $ and becomes leader advantage as the degree of the initial state
entanglement increases. 
\begin{figure}[h]
\begin{center}
\begin{tabular}{ccc}
\vspace{-0.5cm} \includegraphics[scale=0.8]{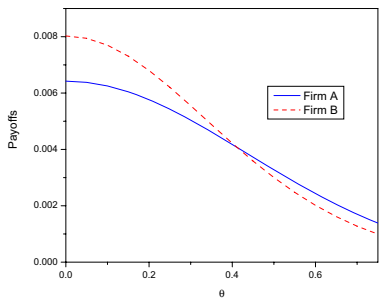}\put(-220,220) &  & 
\end{tabular}%
\end{center}
\caption{\textbf{\ (}color online) The payoffs are plotted at the subgame
perfect Nash equilibrium against the entanglement angle $\protect\theta $.
The values other parameters are chosen as $k=1$, $r=2\protect\pi /9$. The
solid line represents the payoff of firm $A$ and the dotted line represents
the Payoff of firm $B$.}
\label{Fig 4}
\end{figure}

In conclusion, we study the influence of Unruh effect on the payoffs
function of the quantum Stackelberg duopoly. We have shown that the Unruh
effect limits the validation of the subgame perfect Nash equilibrium outcome
to certain range of acceleration of firm $B$'s frame. The acceleration damps
the payoffs function both for unentangled and entangled initial states.
However, the damping is heavy when the initial state is maximally entangled
and the duopoly always benefit the firm that moves first. For an unentangled
initial state, a critical point that correspond to a particular value of the
acceleration exists at which both firms are equally benefitted. For larger
values of acceleration the duopoly becomes a follower advantage. We show
that irrespective of the degree of entanglement in the initial state, the
payoffs function vanish when the acceleration of firm $B$ frame reaches to $%
\pi /4$.

\QTP{Body Math}
{\Huge Acknowledgment}

Salman Khan is thankful to World Federation of Scientists for partially
supporting this work under the National Scholarship Program for Pakistan%
\textbf{.}

{\Huge Figures Captions}\newline
Figure $1$\textbf{. }Rindler spacetime diagram: A uniformly accelerated
observer firm $B$ ($B$) moves on a hyperbola with acceleration $a$ in region 
$I$ and is causally disconnected from region $II$.\newline
Figure $2$\textbf{. (}color online) The payoffs are plotted at the subgame
perfect Nash equilibrium against the acceleration $r$ for unentangled
initial state. The value of $k$ is set to $1$. The solid line represents the
payoff of firm $A$ and the dotted line represents the Payoff of frim $B$.%
\newline
Figure $3$\textbf{.\ (}color online) The payoffs are plotted at the subgame
perfect Nash equilibrium against the acceleration $r$ for maximally
entangled initial state. The value of $k$ is set to $1$. The solid line
represents the payoff of firm $A$ and the dotted line represents the Payoff
of frim $B$.\newline
Figure $4$.\textbf{\ (}color online) The payoffs are plotted at the subgame
perfect Nash equilibrium against the entanglement angle $\theta $. The
values other parameters are chosen as $k=1$,$r=2\pi /9$. The solid line
represents the payoff of firm $A$ and the dotted line represents the Payoff
of frim $B$.\newline

\end{document}